\documentclass[letterpaper,twocolumn,10pt]{article}
\usepackage[hyphens]{url}

\usepackage{usenix-2020-09}

\usepackage[nolist,nohyperlinks]{acronym}
\usepackage[detect-all]{siunitx}
\usepackage{makecell}
\usepackage{amsmath}
\usepackage{cleveref}
\usepackage{tabularx}
\usepackage{booktabs}
\usepackage{float}
\usepackage[ruled]{algorithm2e}
\usepackage[inline]{enumitem}
\usepackage{authblk}

\DeclareSIUnit{\tracepersecond}{trace/\second}
\DeclareSIUnit{\dBm}{dBm}
\DeclareSIUnit{\dBi}{dBi}
\DeclareSIUnit{\bps}{bps}
\DeclareSIUnit{\wpm}{wpm}
\DeclareSIUnit{\cps}{chips/s}

\newcommand{\aref}[1]{\hyperref[#1]{Appendix~\ref{#1}}}
\newcommand{\tlref}[1]{\hyperref[#1]{Table~\ref{#1}}}
\newcommand{\algoref}[1]{\hyperref[#1]{Algorithm~\ref{#1}}}

\usepackage{xspace}
\newcommand{\standard}{IEEE\,802.15.4\xspace}
\newcommand{\standardz}{IEEE\,802.15.4z\xspace}
\newcommand{\standarda}{IEEE\,802.15.4a\xspace}

\usepackage{tikz}
\usetikzlibrary{arrows, calc, positioning, decorations.markings, decorations.pathreplacing, shapes}
\usepackage{pgfplots}
\usepackage{amsmath}
\usepackage{nicefrac}
\usepackage{caption}
\usepackage{subcaption}

\usepackage{listings}
\usepackage{filecontents}

\begin{document}
\date{}

\title{\Large \bf Ghost Peak:\\
  Practical Distance Reduction Attacks Against HRP UWB Ranging}

\author[1,*]{\rm Patrick Leu}
\author[1,*]{\rm Giovanni Camurati}
\author[2]{\rm Alexander Heinrich}
\author[1]{\rm Marc Roeschlin}
\author[1]{\rm Claudio Anliker}
\author[2]{\rm Matthias Hollick}
\author[1]{\rm Srdjan Capkun}
\author[2]{\rm Jiska Classen}
\affil[1]{ETH Zurich}
\affil[2]{TU Darmstadt}
\affil[*]{Authors contributed equally to this research}

\maketitle

\begin{abstract}
We present the first over-the-air attack on \standardz \ac{hrp}
\ac{uwb} distance measurement systems. Specifically, we demonstrate a practical distance reduction attack against pairs of Apple U1 chips (embedded in iPhones and AirTags), as well as against U1 chips inter-operating with NXP and Qorvo UWB chips. These chips have been deployed in a wide range of phones and cars to secure car entry and start and are projected for secure contactless payments, home locks, and contact tracing systems. Our attack operates without any knowledge of cryptographic material, results in distance reductions from \SI{12}{\meter} (actual distance) to \SI{0}{\meter} (spoofed distance) with attack success probabilities of up to \SI{4}{\percent}, and
requires only an inexpensive (USD 65) off-the-shelf device.
Access control can only tolerate sub-second latencies to not inconvenience the user, leaving little margin to perform time-consuming verifications.
These distance reductions bring into question the use of UWB HRP in security-critical applications.
\end{abstract}

\section{Introduction}
\label{sec:intro}
\acl{uwb} chips that measure distance are being massively deployed in smartphones, cars, and other products ~\cite{samsung-gv60, 3db-vw-lrp, vergeAirtag}. Applications range from entry and start systems in cars to mobile payments, contact tracing, spatial awareness, and indoor localization. In addition to enhanced precision compared to more traditional signal strength based ranging, \acs{uwb} aims to provide security against relay and distance reduction attacks~\cite{DBLP:conf/ndss/FrancillonDC11}, which have been used in practice for car thefts and attacks on contactless payments~\cite{SkyNewsCarRelay, ElectrectTeslaRelay, Klee2020Nfcgate}.

The recently adopted \standardz standard~\cite{9179124} aims to address known distance reduction attacks. It introduces two ranging modes: \acf{lrp} and \acf{hrp}. 
Although both modes are used in automotive applications, primarily for \ac{pkes} systems~\cite{3db-vw-lrp, bmw-uwb, nxp-kits, embedded-uwb-adoption}, HRP has seen adoption in Apple iPhones and AirTags, as well as Samsung phones and SmartTags~\cite{nxp-samsung, u1-wired, samsung-smarttag}. 
Despite its standardization and deployment, no public example implementations or standardized algorithms for security-relevant functionality exist. \standardz focuses on message formats without mandating in detail how ranging is done and protected at the endpoints. 

This paper demonstrates the first practical over-the-air distance reduction attack against the \ac{uwb} \standardz \ac{hrp} mode.
Even though the security of \ac{hrp} has been recently studied, these studies were done in simulations~\cite{DBLP:conf/wisec/SinghRZLC21}. We refine existing attacks on \ac{hrp}, introduce a new one, and demonstrate their feasibility in practical settings with Apple U1 (iPhone/AirTag/HomePod), NXP Trimension SR040/ SR150, and Qorvo DWM3000 chips. Our attack enabled a successful distance reduction of up to \SI{12}{\meter} with an overall success rate of \SI{4}{\percent}.
Typically, false acceptance rates are \nicefrac{1}{$2^{20}$} for gate access control and \nicefrac{1}{$2^{48}$} for mobile payments, such that it would take days to years until a fake measurement gets accepted. 

Manufacturers advertise some of the evaluated chips as secure ranging capable~\cite{nxp-factsheet}.
We performed our tests using the configurations that are openly accessible on these chips. Since security algorithms and parameters are not public in the chips that we tested (Apple, NXP, Qorvo), it is hard to determine if these systems can be configured differently and if these alternative configurations would be vulnerable to our or other attacks. Additionally, the past has shown that undisclosed wireless protocols can signify security-by-obscurity solutions\cite{airdrop, magicpairing}. Prior work~\cite{DBLP:conf/wisec/SinghRZLC21} further suggests that making \ac{hrp} ranging both secure and reliable is likely hard. 

The deployment and use of UWB will presumably increase in the future. The FiRa consortium~\cite{firac} has been founded to contribute to the development and widespread adoption of \ac{uwb} technologies in the context of \emph{secured fine ranging and positioning}. The Car Connectivity Consortium recently published Digital Key Release 3.0, enabling \ac{pkes} via UWB in combination with Bluetooth Low Energy \cite{ccc}. At least one car manufacturer has already announced that it will support the iPhone as an access token for \ac{pkes}, citing \ac{uwb} as a ranging mechanism~\cite{bmw-uwb}.  Since \ac{uwb} as an access system is a new protocol, it might take time until malicious actors can fully understand and bypass security checks\cite{180215}. However, systems in cars and other areas related to access control have to be secure for decades after initial deployment.
Therefore, we see this work as another step towards a better understanding of the security of \ac{uwb} \ac{hrp}.

In summary, we make the following contributions:

\setlist[itemize]{itemsep=0em} 
\begin{itemize}
    \item We introduce the first practical distance reduction attack on \standardz \ac{hrp}. Our attack operates in a black-box manner and assumes neither knowledge of cryptographic material shared between the attacked devices nor access to (randomized) ranging message content before messages are transmitted.
    This attack not only validates observations from simulation-based studies of \ac{hrp} but also introduces a novel attack dimension---it selectively varies the power of the injected packet per packet field. The power level is independently adjusted for different fields so that the injected signal is neither perceived as an additional packet nor as jamming the legitimate one. Our attack can therefore also be seen as a type of selective overshadowing. 
    \item We implement our attack on inexpensive (USD 65), commercial off-the-shelf components and demonstrate it on Apple iPhones and AirTags (U1 chip) and on iPhones interoperating with NXP SR040/SR150 and Qorvo DWM3000 UWB chips. We evaluate our attack through a series of experiments and show that the attacker can reduce the measured distances from \SI{12}{\meter} to \SI{0}{\meter} (measured distance). During normal execution, the measurement error is between \SI{10}{\centi\meter} and \SI{20}{\centi\meter}. With a success rate as high as \SI{4}{\percent}, our attack suffices to deceive ranging systems that rely on single HRP measurements.
    
    \item We discuss the implications of our results to different applications and use cases and the applicability of different mitigation techniques in practical settings.

    \item We responsibly disclosed our findings to Apple and NXP, and are in the process of disclosing to Qorvo.
    
\end{itemize}

The rest of the paper is organized as follows. In \autoref{sec:background}, we provide background on \ac{uwb} secure distance measurements. In~\autoref{sec:attack}, we present our attack. We discuss our experimental results in \autoref{sec:evaluation}. Finally, we reflect on the security of \ac{hrp} \ac{uwb} in \autoref{sec:discussion} and compare it to related work in \autoref{sec:related} before concluding in \autoref{sec:conclusion}. 

\section{Background}
\label{sec:background}
In this chapter, we provide the necessary background on \ac{tof} \ac{hrp} \ac{uwb}. We first introduce the concept of \ac{tof} ranging and show how \ac{hrp} uses cross-correlation to determine the \ac{tof} before explaining security considerations behind \ac{hrp}. Finally, we provide a brief overview of available \ac{hrp} chips and products.

\subsection{\ac{uwb} Secure Ranging} \label{sec:uwb_secure_ranging}
The simplicity and practicality of relay attacks on \ac{pkes} systems~\cite{DBLP:conf/ndss/FrancillonDC11, SkyNewsCarRelay, ElectrectTeslaRelay} urged a paradigm shift in secure ranging. Utilizing a signal's \ac{tof} is promising since a relay can only increase the \ac{tof} and, thus, the measured distance. However, research in this field has shown that such systems can still be vulnerable to more sophisticated attacks, such as Cicada ~\cite{5616900} or \ac{edlc}~\cite{DBLP:journals/twc/PoturalskiFPHB11}.

\ac{uwb} aims to implement secure ranging, including physical-layer security~\cite{fira-phy}.
\standard proposes two modes for UWB ranging named \ac{lrp} and \ac{hrp}. They are both subject to stringent power limitations, as their channels overlap with frequency bands used by existing technologies, such as \mbox{Wi-Fi} or cellular networks. While LRP approaches the power limit by using fewer but stronger pulses (each individually 'visible' to the  receiver), HRP relies on a larger number of weaker pulses (which cannot be individually decoded in most environments by the receiver). This difference in design has consequences; while the security of LRP is easy to demonstrate, the resilience of HRP against reduction attacks is an open research question. Recent in-simulation analysis of HRP has shown that HRP might be hard to configure to be both performant and secure~\cite{DBLP:conf/wisec/SinghRZLC21}. 

\subsubsection{\acl{twr}}

\input{figures/hrp_ranging.tikz}

The \standard standard defines three different ranging and localization methods, namely \acf{ss-twr}, \acf{ds-twr}, and \ac{tdoa}; our work focuses on \ac{ss-twr} and \ac{ds-twr}. 
\ac{tdoa} mode is not in the scope of our work since, to the best of our know\-ledge, none of the currently available \ac{uwb} transceiver chips support this method.

\textbf{\ac{ss-twr}} is depicted in \autoref{fig:sfiga}, which shows how \ac{tof} for the distance calculation can be determined by subtracting $T_{reply}$, the processing time of the responder, from $T_{round}$, the total round trip time measured by the initiator. Dividing the result by two yields an estimation of the propagation delay $\hat{T}_{prop}$, or the \ac{tof} required by the signal to cover one way. However, this result may be affected by a possible clock frequency offset between initiator and responder. If the initiator can measure this offset, it can compensate for it and improve the measurement.

\textbf{\ac{ds-twr}}, as shown in \autoref{fig:sfigb}, mitigates the clock offset by transmitting more messages. \ac{ds-twr} comprises two \ac{ss-twr} exchanges in opposite directions. $T_{reply}$ and $T_{round}$ are measured with both devices/clocks, significantly reducing errors induced by clock offset and drift. 
\ac{ds-twr} is optimized by simultaneously using the response message of the first exchange as the request message of the second, thus reducing the procedure to three ranging messages. The derivation of the propagation time formula can be found in~\cite{DBLP:conf/wpnc/NeirynckLM16}. 

\subsubsection{Receiver Design and Cross-Correlation}\label{subsubsec:receiverdesign}
Most RF communication technologies rely on cross-correlation to detect the presence of an incoming message. In UWB, the receiver constantly scans the acquired signal for a static (pre-negotiated) preamble using a local template.
The received signal is digitized and recorded as I/Q samples fed into a correlator. If the output exceeds the level for noise by a certain amount, the receiver concludes that a packet must be present and analyzes the signal further.


\tikzset{>=latex}
\definecolor{darkblue}{RGB}{0,73,200}
\definecolor{darkred}{RGB}{212,0,16}
\definecolor{darkgreen}{RGB}{11,196,1}

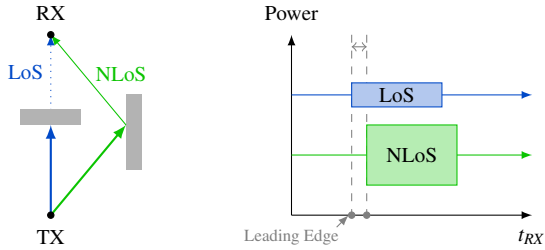
\begin{figure}[!b]

\centering

	\center
	\begin{tikzpicture}[minimum height=0.7cm, scale=0.8, every node/.style={scale=0.8}, node distance=0.9cm]

	\draw[thick, ->, darkblue] (0, -3) -- (0, -1.5);
	\path[->, darkblue, dotted] (0, -1.25) edge node[left] {LoS} (0, 0);
	\fill[fill=darkgray!40] (-0.5,-1.5) rectangle node (a1) {} ++(1, 0.25); 
	
	\fill[fill=darkgray!40] (1.25,-2.25) rectangle node (a1) {} ++(0.25, 1.25); 
	\draw[thick, ->, darkgreen] (0, -3) -- (1.25, -1.5);
	\path[->, darkgreen] (1.25, -1.5) edge node[right,yshift=0.1cm] {NLoS} (0, 0);

	\filldraw[black] (0,0) circle (1.5pt);
	\node[above] at (0,0) {RX};
	\filldraw[black] (0,-3) circle (1.5pt);
	\node[below] at (0,-3) {TX};
	
    \begin{scope}[shift={(4,0)}]
    
    \draw[->] (0, -3) -- (0, 0);
    \node[align=center, above] at (0, 0) {Power};
    \draw[->] (0, -3) -- (4, -3);
    \node[align=center, below] at (4, -3) {$t_{RX}$};

	\draw[-, gray, dashed] (1, 0) -- (1, -3);
	\draw[-, gray, dashed] (1.25, 0) -- (1.25, -3);
	\filldraw[gray] (1,-3) circle (1.5pt);
	\filldraw[gray] (1.25,-3) circle (1.5pt);
	\path[<->, gray, >=to] (1, -0.2) edge (1.25, -0.2);
	\node[gray, below] at (0, -3) {\footnotesize{Leading Edge}};
	\draw[gray, ->] (0.75, -3.2) -- (0.95, -3);
	
	\draw[->,darkblue] (0, -1) -- (4, -1);
	\draw[->,darkgreen] (0, -2) -- (4, -2);
	\filldraw[fill=darkgreen!30, draw=darkgreen] (1.25, -2.5) rectangle node (r) {\textcolor{black}{NLoS}} ++(1.5, 1);
	\filldraw[fill=darkblue!30, draw=darkblue] (1, -1.2) rectangle node (r) {\textcolor{black}{LoS}} ++(1.5, 0.4);

	\end{scope}

\end{tikzpicture}

\caption{\label{fig:leading-edge} In a \ac{nlos} scenario, the receiver needs to detect the arrival time of the early \ac{los} copy (leading edge).}

\end{figure} 

\tikzset{>=latex}
\definecolor{darkblue}{RGB}{0,73,200}
\definecolor{darkred}{RGB}{212,0,16}
\definecolor{darkgreen}{RGB}{11,196,1}

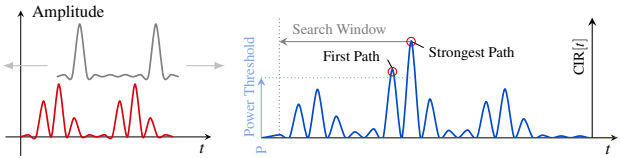
\begin{figure}[!b]
\centering

	\center
	\begin{tikzpicture}[minimum height=0.7cm, scale=0.8, every node/.style={scale=0.8}, node distance=0.9cm]

	\begin{axis}[axis lines=middle,
		ticks=none,
		ymin=-0.9,
		ymax=6,
		xmin=-1,
		xmax=12.5,
		ylabel={Amplitude},
		y label style={at={(0.1,1.1)}},
		xlabel={$t$},
		x label style={at={(1,-0.07)}},
		height=4cm,
		width=5cm]
	\addplot[darkred, smooth,thick] coordinates {(0,0) (0.5,0.1) (1,0) (1.5,1.7) (2,0.0) (2.5,2.5) (3.0,0.0) (3.5,0.9) (4, 0.0) (4.5, 0.1) (5,0) (5.5, 0.1) (6,0) (6.5, 1.7) (7,0) (7.5,2.5) (8,0) (8.5, 0.9) (9, 0) (9.5, 0.1) (10, 0)
	};
	\addplot[gray, smooth,thick,xshift=0.6cm, yshift=1cm] coordinates {(0,0) (0.5,0.1) (1,0) (1.5,2.5) (2,0.0) (2.5,0.1) (3.0,0.0) (3.5,0.1) (4, 0.0) (4.5, 0.1) (5,0) (5.5, 0.1) (6,0) (6.5, 2.5) (7,0) (7.5,0.1) (8,0)};
	\end{axis}
	
	\draw[<-, gray!50] (-0.05,1.5) -- (0.7,1.5);
	\draw[->, gray!50] (3.0,1.5) -- (3.65,1.5);


	\begin{scope}[shift={(4.25,0.3)}]
	\begin{axis}[axis lines=middle,
	    axis y line=none, 
		ticks=none,
		ymin=0.0,
		ymax=3.5,
		xmin=-7,
		xmax=12,
		ylabel={$\mathrm{CIR}[t]$}, 
		y label style={at={(0.76,1.05)}, rotate=-90},
		xlabel={$t$},
		x label style={at={(1,-0.22)}},
		height=3.75cm,
		width=7.5cm]
	\addplot[darkblue, smooth,thick] coordinates {(-12, 0.0) (-10,0.0) (-9,0.0) (-8, 0) (-7, 0.0) (-6,0.08) (-5.5,0) (-5, 0.9) (-4.5, 0) (-4, 1.3) (-3.5, 0) (-3,0.58) (-2.5,0) (-2,0.16) (-1.5,0) (-1.0,0.24) (-0.5,0) (0,1.84) (0.5, 0.0) (1.0, 2.6) (1.5,0) (2.0,1.04) (2.5, 0) (3.0, 0.2) (3.5, 0) (4, 0.21) (4.5, 0) (5, 0.97) (5.5, 0) (6, 1.31) (6.5, 0) (7, 0.51) (7.5,0) (8,0.07) (8.5,0) (9,0.05) (9.5,0) (10,0.05) (10.5,0) (11,0) (11.5,0) (12,0)};
	\end{axis}
	
	\path[-stealth, black] (5.5,0) edge node[rotate=90,yshift=0.2cm, xshift=0.35cm]{\scriptsize $\mathrm{CIR}[t]$} (5.5,2);

	\path[stealth-, gray] (0.3,1.6) edge node[above,yshift=-0.15cm,xshift=-0.1cm] {\scriptsize Search Window} (2.5,1.6);
	\draw[densely dotted, gray] (0.3,0) -- (0.3,2);
	
	\node[text=darkblue!50, rotate=90, xshift=-0.2cm] at (0,0) {\scriptsize P};
	\draw[densely dotted, darkblue!50] (0,1.0) -- (2.5,1.0);
	\path[-stealth, darkblue!50] (0,0) edge node[rotate=90,yshift=0.2cm, xshift=0.35cm]{\scriptsize Power Threshold} (0,1);
	
	\node[] at (3.5, 1.4) {\scriptsize Strongest Path};
	\draw[] (2.5, 1.6) -- (2.7, 1.5);
	\draw[darkred] (2.49, 1.6) circle (2pt);
	\node[above, align=center] at (1.5, 1) {\scriptsize First Path};
	\draw[] (2.1, 1.3) -- (2.18, 1.15);
	\draw[darkred] (2.18, 1.1) circle (2pt);
	
	\end{scope}

	\end{tikzpicture}

\caption{\label{fig:cross-correlation} The \acs{cir} is calculated based on the received signal and a local template of the expected signal.
A correlation peak indicates high similarity. However, with multi-path effects, there are multiple peaks.
The receiver identifies the first \ac{los} path, e.g., by searching back in time from the strongest peak.
}

\end{figure}

In practice, this process has to be optimized to cope with channel distortions, most notably multi-path fading.
During transit, objects in the vicinity reflect the signal, which creates copies of the signal that are slightly delayed in time, as shown in \autoref{fig:leading-edge}. Those copies are superimposed onto the original signal, causing constructive or destructive interference. 
Therefore, a momentary output of the correlation is non-conclusive, and instead, the \acf{cir}, i.e., the correlation output over time, must be inspected. The CIR greatly supports the search for a known template in the received signal and can determine the precise arrival time of a packet.
The \ac{cir} can be estimated as follows:
\[ \mathrm{CIR}[t] = (g_{\mathrm{loc}} * s)[t] = \sum_{m=0}^{|g_{\mathrm{loc}}|-1}{\overline{g_{\mathrm{loc}}[m]} \cdot s[m+t]}\]
where $s[\cdot ]$ is the complex and time-discrete received signal and $g_{\mathrm{loc}}[\cdot ]$ is the template of the expected signal.

As shown in \autoref{fig:cross-correlation}, \ac{hrp} \ac{uwb} ranging relies heavily on cross-correlation, to detect and determine the arrival times of preamble and a \ac{sts}. We explain STS in the next section.
A UWB receiver cross-correlates the incoming signal with a template (e.g., a known sequence) for the preamble and, if present, also with a known template for the \ac{sts}.
High correlation values imply similarities between the template and the received signal. However, the CIR only shows a single distinct peak in perfect conditions. Due to multi-path, the CIR often shows a profile containing several peaks, and it is not straightforward to identify the \emph{first} peak/path that reflects the actual physical distance. Constructive and destructive interference can lead to a CIR where the first path emerges as a peak with an amplitude significantly below the maximal value. Channel and receiver noise make the search for the first path and thus the correct distance even more challenging.

\subsubsection{High-Rate Pulse Repetition (\ac{hrp})}\label{subsubsec:hrp}
HRP mode of \standard uses a high pulse repetition frequency of \SI{64}{\mega\hertz}. The spacing between pulses is narrow and, to meet stringent restrictions on power spectral density (\num{-41.3}\,dBm/MHz)~\cite{uwb-regulations}, the power per pulse is low, in the order of \num{-80},\ instantaneous dBm (at the antenna port).
The information elements of a packet are either encoded with \ac{bpm} using \ac{bpsk} or just BPSK symbols.
In BPM-BPSK, a symbol can encode two bits by varying the position of the burst and the polarity of the pulses,
while in BPSK, a positive polarity pulse encodes a bit of value zero, and a negative polarity pulse (\SI{180}{\degree} phase shift) encodes a bit of value one. 
Most UWB channels are \SI{499.2}{\mega\hertz} wide, which is the bandwidth used by all our tested devices. At \SI{499.2}{\mega\hertz}, the duration of a pulse is in the order of \SI{2}{\nano\second}.


\tikzset{>=latex}

\begin{figure}[b]

	\center
	\begin{tikzpicture}[minimum height=0.55cm, scale=0.8, every node/.style={scale=0.8}, node distance=0.7cm]

	\filldraw[fill=darkgray!20, draw=darkgray](0,-0.25) rectangle node (a1) {Preamble} ++(2, 0.7);
	\filldraw[fill=darkgray!20, draw=darkgray](2,-0.25) rectangle node (a1) {SFD} ++(1, 0.7);
	\filldraw[fill=darkgray!20, draw=darkgray](3,-0.25) rectangle node (a1) {STS} ++(1, 0.7);
	\filldraw[fill=darkgray!20, draw=darkgray](4,-0.25) rectangle node (a1) {PHR} ++(1, 0.7);
	\filldraw[fill=darkgray!8, draw=darkgray, dashed](5,-0.25) rectangle node (a1) {Data payload} ++(4, 0.7);
	
	

	\end{tikzpicture}

\caption{\label{fig:hrp-packet} Example format of an \ac{hrp} packet~\cite{9179124}. The lengths of the different parts and their order depend on the configuration.}

\end{figure}
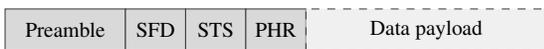

\textbf{\ac{hrp} PHY Packet} 
It is essential for the attacks described in this paper to understand the \ac{hrp} packet construction and pulse sequence.
\autoref{fig:hrp-packet} shows the different segments that constitute an \ac{hrp} ranging message using an \ac{sts}. 
 The preamble of the packet is used to detect the presence of a ranging message. The \ac{sts} contains a pseudo-random bit sequence for security purposes, and the data segment may be used to transmit additional information. The \ac{sfd} should be taken as a reference to calculate the propagation delay, and the PHR carries the physical header of the packet. We refer interested readers to the official release of the \standard standard for a more detailed description of the PHY~\cite{6012487}.

\textbf{Scrambled Timestamp Sequence (\ac{sts})}
The preamble is a pre-defined and static sequence of pulses representing \num{-1}, \num{0}, and \num{+1},
modulated using a ternary code, i.e., positive, negative and no pulse. In contrast, the \ac{sts} consists of BPSK-modulated pulses representing \num{-1} and \num{+1}. The bit sequence in the \ac{sts} is the output of a pseudo-random generator and derived as outlined in \autoref{fig:sts-generation}.
The ranging devices need to agree on a 128-bit key, e.g., by using an out-of-band channel, before the UWB ranging operation can commence.
A \emph{fresh} STS is generated for every ranging message, as the $STS\ V\ Counter$ increases with every packet.
The ranging devices know in advance what bit sequence to expect in place of the STS, and they can create a local template to detect the incoming STS using cross-correlation as described in~\autoref{subsubsec:receiverdesign}.
Since the STS contents cannot be predicted, it is theoretically impossible for an external source to emit a signal that arrives at the targeted device earlier in time and still contains the legitimate STS; only the legitimate device knows what to send. 
As a result, ranging devices can base the \ac{toa} of the packet on the arrival time of the STS and thereby guarantee that no external adversary reduces the measured distance by advancing the received signal in time.
Moreover, it is also impossible to react to isolated pulses and send those earlier in time since the BPSK pulses as part of the STS are only about \SI{2}{\nano\second} long. An adversary can not acquire the polarity of a single pulse in sufficient time, which makes any replay or \ac{edlc} attack physically impossible. In any case, advancing pulses would only yield a \SI{2}{\nano\second} reduction at the maximum, translating to less than \SI{60}{\centi\meter} in distance.

\textbf{Channel Distortion and Multi-Path Fading}
UWB ranging packets are subject to channel noise and multi-path fading,
rendering the (direct) demodulation of single pulses of the STS intricate and in some channel conditions impossible. At a \SI{64}{\mega\hertz} pulse repetition frequency, the pulse spacing is in the order of \SI{16}{\nano\second}, which is above the channel delay spread. As a consequence, inter-pulse interference and multi-path fading effects make separate pulses unrecognizable.
To work around this, HRP detects STS by cross-correlating the received signal with the expected STS, similar to preamble detection. 
Although cross-correlation is a powerful tool to determine the presence of the STS, the computed \acf{cir} often shows a profile that contains multiple correlation peaks, and pinpointing the exact arrival time remains challenging. The CIR is a superposition of cross-correlation side peaks and weak early path correlation peaks.
\autoref{fig:cross-correlation} shows two pulses after reception (in red) and the template used by the receiver (in grey). The resulting CIR (in blue) exhibits multiple peaks. The highest peak does not necessarily correspond to the \ac{los} path of the signal.
Even \emph{before} the strongest correlation value, any \ac{hrp} receiver must check for additional peaks within a specific time window. 
Such a peak might suggest an earlier but weaker copy of the signal, which belongs to a shorter path. By using this path as a reference, the receiver can compute a more accurate ranging result. Details on how the time of arrival of the STS is determined are not specified in the standard for HRP. At the time of writing, the exact procedure remains protected intellectual property for all commercially available HRP transceiver chips we have evaluated.

\subsubsection{Ideal versus Real Security Guarantees}
If every pulse contained in the STS would be demodulated, absent of noise and channel effects, the receiver could verify every single bit in the sequence.
However, in \ac{hrp} \ac{uwb}, the \num{4096}\,bit long \ac{sts} does not result in \num{4096}\,verifiable bits. First, the entropy of the key used for AES in counter mode is only \num{128}\,bits, see \autoref{fig:sts-generation}.
Second, since the \ac{sts} is verified by correlation instead of single pulse demodulation, any security guarantee is given by the significance level of the early peak compared to the overall cross-correlation profile. Non-ideal cross-correlation properties of random sequences, such as the \ac{sts}, can cause side-lobes in the correlation and play a minor role.

A bit-wise \ac{sts} comparison instead of cross-correlation would have to allow for transmission errors, which naturally happen in \ac{nlos} scenarios.
\standardz does not specify whether \ac{sts} should be compared bit-wise after correlation operation.
Even if a vendor implements additional checks, they need to account for bit flips and choose a threshold that significantly impact on the security provided by the \ac{sts}.


\tikzset{>=latex}
\definecolor{darkblue}{RGB}{0,73,200}

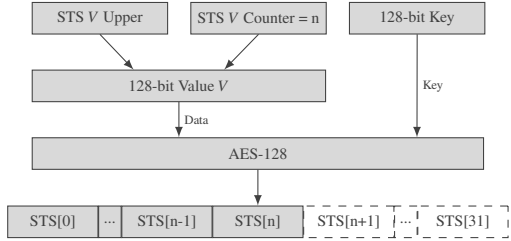
\begin{figure}[t]

	\center
	\begin{tikzpicture}[minimum height=0.7cm, scale=0.6, every node/.style={scale=0.6}, node distance=0.9cm]

	\node[fill=darkgray!20, text=darkgray, draw=darkgray, minimum width=2cm,  anchor=west] (first) {STS[0]};
	\node[fill=darkgray!20, text=darkgray, draw=darkgray, minimum width=0.5cm, right=of first, xshift=-1.5 cm] (second) {...};
	\node[fill=darkgray!20, text=darkgray, draw=darkgray, minimum width=2cm,  right=of second, xshift=-1.5 cm] (lastblock) {STS[n-1]};
	\node[fill=darkgray!20, text=darkgray, draw=darkgray, minimum width=2cm, right=of lastblock, xshift=-1.5 cm] (prng) {STS[n]};
	\node[text=darkgray, draw=darkgray, dashed, minimum width=2cm, right=of prng, xshift=-1.5 cm] (nextblock) {STS[n+1]};
	\node[text=darkgray, draw=darkgray, dashed, minimum width=0.5cm, right=of nextblock, xshift=-1.5 cm] (secondlast) {...};
	\node[text=darkgray, draw=darkgray, dashed, minimum width=2.0cm, right=of secondlast, xshift=-1.5 cm] (last) {STS[31]};
	\node[fill=darkgray!20, text=darkgray, draw=darkgray, minimum width=10cm, above=of prng, yshift=-0.7cm] (aes) {AES-128};	
	\node[fill=darkgray!20, text=darkgray, draw=darkgray, minimum width=6.5cm, above=of aes.west, anchor=west] (v) {128-bit Value $V$};	
	\node[fill=darkgray!20, text=darkgray, draw=darkgray, minimum width=3cm, xshift=0cm, above=of v.west, anchor=west] (iv) {STS $V$ Upper};	
	\node[fill=darkgray!20, text=darkgray, draw=darkgray, minimum width=3cm, xshift=-1cm, right=of iv.east, anchor=west] (ctr2) {STS $V$ Counter = n};	
	\node[fill=darkgray!20, text=darkgray, draw=darkgray, minimum width=3cm, xshift=-1cm, right=of ctr2.east, anchor=west] (keys) {128-bit Key};	
	
	\draw[->,darkgray] (aes.south) -- (prng.north);
	\path[->,darkgray] (v.south) edge node[anchor=west] {\footnotesize{Data}} ([xshift=-1.75cm] aes.north);
	\path[->,darkgray] (keys.south) edge node[anchor=west] {\footnotesize{Key}} ([xshift=3.525cm] aes.north);
	\draw[->,darkgray] (iv.south) -- ([xshift=-1cm] v.north);
	\draw[->,darkgray] (ctr2.south) -- ([xshift=1cm] v.north);

	\end{tikzpicture}

\caption{\label{fig:sts-generation}
Cryptographically secure \ac{sts} generation with AES in counter mode. Each iteration results in a random 128-bit block. The STS $V$ Counter is incremented for every iteration. The entire STS comprises 32 blocks, or 4096 bits~\cite{9179124}.}

\end{figure}

\subsection{Commercial HRP UWB Chips}

As of now, only a few vendors offer HRP transceiver chips,
despite the fact that HRP-based location and tracking tags have entered the consumer market at scale~\cite{vergeAirtag} and automotive manufacturers are planning to release cars featuring \ac{pkes} systems built on top of HRP chips, such as the BMW iX and the Genesis GV60 models~\cite{bmw-uwb, samsung-gv60}. The FiRa consortium considers HRP viable for both consumer-grade and security-critical applications alike~\cite{fira-phy}.

\textbf{Apple} has a diverse \ac{uwb} software and hardware stack. Different versions of the Apple U1 chip have been released in recent products, such as the iPhone (since iPhone 11), the HomePod mini, the Apple Watch (since Series 6), and even the USD 30 AirTag.
On the iPhone, Apple integrated \ac{uwb} into AirDrop with iOS 13~\cite{airdrop-uwb}, using \ac{aoa} measurements to simplify the location of devices and enhance user experience. With iOS 14, they introduced the Nearby Interaction framework, exposing a selected set of \ac{uwb}-based ranging functionality to application developers ~\cite{apple-nearbyinteraction}. A compatibility mode for third-party accessory support has been available since the release of iOS 15~\cite{apple-nearby-accessory}.
However, details about the compatibility mode configuration parameters are only available to \ac{mfi} program members.

\textbf{NXP} advertises their Trimension chip series for secure ranging and precise positioning~\cite{nxp-factsheet}.
Development kits exist for the SR150 and SR040~\cite{nxp-kits}. Our analysis showed that several Samsung products, for example, the SmartTag+ and phones starting from Samsung Note20 Ultra ~\cite{nxp-samsung}, contain NXP chips to enable ranging and improve Point to Share ~\cite{samsung-p2s} data transfers. Examples for cars that comprise NXP chips are upcoming BMW and VW models ~\cite{nxp-bmw, nxp-vw}, whereas VW seems to incorporate \ac{lrp} chips for PKES use cases \cite{3db-vw-lrp}.

\textbf{Qorvo}, also known as Decawave before their acquisition~\cite{qorvo-acquisition},
manufactures the DW3000 chip series.
These chips are interoperable with the Apple U1 chip~\cite{DW3110Qorvo}. Nevertheless, to the best of our knowledge, there are no commercially available products that use the DW3000 series and are compatible with Samsung or Apple consumer devices.
Qorvo also offers two development kits: DWM3000EVB, an Arduino-based development board~\cite{DWM3000EVBQorvo}, and DWM3001CDK, an integrated board that contains an nRF52833 with Bluetooth 5.2~\cite{DWM3001CDKQorvo}.
Both of them can be programmed using their SDK. 

\section{A Practical Distance-Reduction Attack}

\label{sec:attack}
In the following, we explain our attacker model, the theoretical working
principle of our attack, including boundaries of distance reduction,
and the attack algorithm and setup.

\subsection{Attacker Model and Attack Overview}
\label{ssec:attackermodel}
\input{figures/overview.tikz}

\tikzset{>=latex}

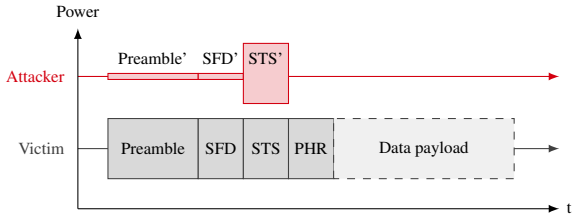
\begin{figure}[t]

	\center
	\begin{tikzpicture}[minimum height=0.55cm, scale=0.8, every node/.style={scale=0.8, font={\footnotesize}}, node distance=0.7cm]
	
    \node[left,darkred] at (-0.1,0.2) {Attacker};
    \draw[->,darkred] (0, 0.15+0.05) -- (8, 0.15+0.05);
	\filldraw[fill=darkred!20, draw=darkred](0.5,0.15) rectangle node (a1) {} ++(1.5, 0.1);
	\filldraw[fill=darkred!20, draw=darkred](2,0.15) rectangle node (a1) {} ++(0.75, 0.1);
	\filldraw[fill=darkred!20, draw=darkred](2.75,-0.25) rectangle node (a1) {} ++(0.75, 1);
    \node[align=center] at (0.5+0.75, 0.5) {Preamble'};
    \node[align=center] at (2+0.375, 0.5) {SFD'};
    \node[align=center] at (2.75+0.375, 0.5) {STS'};
	
    \node[left,darkgray] at (-0.1,-1) {Victim};
    \draw[->,darkgray] (0, -1) -- (8, -1);
	\filldraw[fill=darkgray!20, draw=darkgray](0.5,-1.5) rectangle node (a1) {Preamble} ++(1.5, 1);
	\filldraw[fill=darkgray!20, draw=darkgray](2,-1.5) rectangle node (a1) {SFD} ++(0.75, 1);
	\filldraw[fill=darkgray!20, draw=darkgray](2.75,-1.5) rectangle node (a1) {STS} ++(0.75, 1);
	\filldraw[fill=darkgray!20, draw=darkgray](3.5,-1.5) rectangle node (a1) {PHR} ++(0.75, 1);
	\filldraw[fill=darkgray!8, draw=darkgray, dashed](4.25,-1.5) rectangle node (a1) {Data payload} ++(3, 1);

    \draw[->] (0, -2) -- (0, 1);
    \node[align=center, above] at (0, 1) {Power};
    \draw[->] (0, -2) -- (8, -2);
    \node[align=center, right] at (8, -2) {t};

	\end{tikzpicture}

\caption{\label{fig:overshadow} The attacker overshadows the legitimate signal (gray) with a carefully crafted attack packet (red). The attacker's STS' is random, without knowing the victim's STS.}

\end{figure}
We consider an attacker that is trying to reduce the distance measured between two \ac{hrp} \ac{uwb} devices. E.g., an attacker trying to unlock and start a car by tricking it into believing that the legitimate owner's car keyfob is near. 

We consider a \emph{black-box} attacker with the following limitations.
The attacker has no access to any secrets shared between victim devices and cannot predict message field contents that are assumed to be unpredictable in HRP UWB; i.e., the attacker cannot predict the \acf{sts}.
Unable to guess the \ac{sts}, our attacker cannot simply send a valid packet to advance the
message time of arrival and therefore reduce the distance. 
The attacker can place its devices in physical proximity to one of the victim devices but has no physical access and cannot tamper with these devices.
The attacker can receive and inject signals on the wireless communication channel. Specifically, they can craft and transmit UWB messages based on the \ac{hrp} standard. 

We illustrate our attack in \autoref{fig:overview} and \autoref{fig:overshadow}.
During an initial attack phase, the attacker device behaves like a \ac{hrp} \ac{uwb} packet analyzer and resolves the sequences of packets exchanged by the victim devices. 
Once the ranging sequence and its timings have been identified, the attacker device reactively overshadows
selected packet components, as shown in \autoref{fig:overview2}.
The additional noise caused at the receiver by overshadowing is
misclassified as an early copy of the legitimate signal, causing
distance reduction. 

Injected overshadowing signals follow the structure of an HRP PHY packet but are crafted such that different packet fields are transmitted at different power levels. 
The attacker needs to synchronize to packet transmissions. Packets are sent every few \SI{}{\milli\second}, depending on the ranging implementation,
and the attacker's synchronization accuracy only needs to be in the order of \SI{}{\micro\second}, despite attacking a protocol that measures timings
with \SI{}{\pico\second} resolution~\cite{DW3000USERMANUAL}.

\begin{table*}[!t]
\caption{Comparison with previous work exploiting leading edge detection.}
\label{table:comparison-with-previous-work}
\centering
\footnotesize
\renewcommand{\arraystretch}{1.3}
\begin{tabular}{r|llllll}
\toprule
& \textbf{\acs{uwb} Standard} &
\textbf{Security} &
\textbf{Victim} &
\textbf{Attacked Field} &
\textbf{Attacker} &
\textbf{Attack signal} \\
\midrule
\textbf{\textit{This Paper}} &
\standardz &
STS &
Apple U1 &
STS &
Off-the-Shelf &
Weak Preamble + Strong Random STS\\

Adaptive~\cite{DBLP:conf/wisec/SinghRZLC21} &
\standardz &
STS &
Simulation &
STS &
Simulation & 
Pulses at lower rate for shorter time\\

Cicada++~\cite{DBLP:conf/wisec/SinghRZLC21} &
\standardz &
STS &
Simulation &
STS &
Simulation & 
Pulses at lower rate\\
Cicada~\cite{5616900} &
\standarda &
None &
Simulation &
Preamble &
Simulation &
Pulses \\
\bottomrule
\end{tabular}
\end{table*}

Our attack can be implemented and executed using a simple and inexpensive off-the-shelf \ac{hrp} \ac{uwb} device. Therefore, no complex laboratory equipment is needed, making the attack practical and easy to implement.
\autoref{fig:overshadow} shows an injected packet aligned with a legitimate
packet. The injected packet is composed of Preamble, \ac{sfd}, and 
\ac{sts}, where the \ac{sts} is randomly generated without any knowledge of the legitimate \ac{sts}. The power of each field is independently adjusted
to obtain optimal results, as explained in more detail in
\autoref{sec:working-principle}.

In many practical cases, \ac{hrp} \ac{uwb} devices use \ac{ds-twr} and
possibly exchange additional synchronization or data packets. This information can also be exchanged out-of-band (e.g., using Bluetooth~\cite{ccc, u1-bh}, NFC, UHF). However, this pre-negotiation does not impact the attack, which only targets the time of arrival of packets in the ranging sequence~\cite{DBLP:conf/sp/LeuSRPC20}.
As shown in \autoref{fig:delay}, the attack can be easily generalized. By configuring the delay of reaction after the reception of the
first packet, the attacker can attack any desired packet in the sequence.
In the case of \ac{ds-twr}, this can be leveraged to select the device to attack.
Alternatively, an attacker could also use two
devices to attack both ends simultaneously, increasing the chances
of success.

In summary, an attacker needs to configure which packet in the ranging sequence to attack (by selecting the delay from the reception of the first packet) and the power of the Preamble, \ac{sfd}, and \ac{sts} to inject.
In \autoref{sec:working-principle}, we will explain why and how these
parameters affect the distance measurement.

\subsection{Working Principle}
\label{sec:working-principle}

\subsubsection{Secure Leading Edge Detection}
Accurate timestamps require detecting the earliest copy of the received signal, also called \emph{leading edge detection}. 
In the following, we explain the challenge of leading
edge detection and describe how our attack selectively attacks specific fields of targeted packets in a ranging sequence by overshadowing the contents.

In a realistic environment with obstacles and reflections, the receiver will likely be presented with multiple copies of the transmitted signal, arriving with different power from different paths. 
In \ac{hrp}, the problem is exacerbated because these delays might cause self-interference among pulses that are spaced only by \SI{16}{\nano\second} (high repetition frequency of \SI{64}{\mega\hertz}), see \autoref{subsubsec:hrp}.
For \acl{tof} measurements used in \acl{twr}, the receiver must find the earliest copy, corresponding to the shortest path (\acl{los}). When receiver and transmitter are not in \ac{los}, the \ac{los} copy is likely to arrive at lower power than other \ac{nlos} reflections, as previously shown in \autoref{fig:leading-edge}.
When looking for the leading edge copy, any algorithm or implementation must decide whether it faces noise or a very low power early copy of the signal, which is challenging.

Suppose an attacker is able to inject noise that looks reasonably similar to a legitimate low-power copy to the reception algorithm. In that case, it might trick the receiver into accepting it as the leading edge, causing a distance reduction.
This attack has been first prop,osed for \standarda in~\cite{DBLP:journals/twc/PoturalskiFPHB11}. In \standarda there is no \ac{sts} and the attacker can inject a \ac{uwb} pulse to attack the preamble. A recent study~\cite{DBLP:conf/wisec/SinghRZLC21} has made the hypothesis, confirmed by simulation, that variations of the Cicada attack can be used to attack the \ac{sts} in \standardz by injecting \ac{hrp} pulses. Since \ac{hrp} \ac{uwb} reception algorithms are not publicly known, simulations are based on three main assumptions: (i) arrival time and quality of the \ac{sts} are computed via time-domain cross-correlation, (ii) the leading edge is found by looking for a smaller correlation peak in a limited back-search window before the strongest peak, and (iii) thresholds are set to evaluate the significance of correlation compared to noise. Simulations in~\cite{DBLP:conf/wisec/SinghRZLC21} highlight that, given a reception algorithm, there is a fundamental trade-off between security and performance: lax thresholds are necessary to accept legitimate early copies in challenging multi-path environments, but this increases the chance of accepting attacker-induced noise. 

In this paper, we take the opposite approach. Instead of hypothesizing a certain algorithm and design choice and studying it in simulation, we empirically analyze the behavior of the unknown algorithms deployed in real products (Apple U1) when subject to signal injection.
Because of their closed-source nature, we do not know most of the design choices. For example, we are not aware whether they implement time-domain cross-correlation or take a frequency-domain approach, how they estimate the noise floor, how they define, and configure thresholds and whether such thresholds are dynamically adjusted to the environment.

The only assumption we make when developing our attack is that the receiver is able to work in \ac{nlos} conditions, which we were able to confirm empirically. We then chose to transmit signals crafted from standard packets, to maximize the probability of generating noise that is misclassified for a legitimate copy and to make the attack practical to implement.
Instead of injecting fine-grained aligned pulses at different power and repetition frequencies, we observe how the fields of standard packets affect reception. We adapt the structure of the packet and the power level of the fields to maximize the chances of reduction (by injecting \ac{sts} pulses) while avoiding jamming and other errors. In general, differently from previous work, our attack handles many of those challenges due to the fact that it operates on real sequences of packets used in real exchanges.

It is worth noticing that the attacker does not have direct control over the amount of distance reduction. A method to gain partial control has been proposed in simulation in~\cite{DBLP:conf/wisec/SinghRZLC21}. However, it requires to delay the legitimate copy, emulate the leading edge detection algorithm at reception, analyze its output in real-time, and interrupt the injection when the desired result is obtained. For these reasons, it is hard to implement in practice, in particular with off-the-shelf devices. As an alternative, in \autoref{subsec:packet-selection} we show how the choice of the victim packet(s) in a sequence can affect the distribution of reduction, and in \autoref{sec:evaluation} we empirically analyze it.
In \autoref{table:comparison-with-previous-work} we compare previous work on leading edge detection with our approach.
Nevertheless, in our threat model the attacker is not interested in controlling the reduction but in causing practical distance reductions that will trick the victim into believing the legitimate user is close enough to grant access.

\input{figures/generalization.tikz}

\subsubsection{Selective Overshadowing to Avoid Jamming}
An attack against leading edge detection can be successful in practice only if the injection of the attack signal does not accidentally produce other errors that invalidate a ranging sequence. To achieve this goal, our attack carefully crafts the timing, format, and power level of the attack signal. The attacker's transmission is not continuous but reactive. As opposed to the continuous transmission of Gaussian noise or \ac{uwb} pulses, a reactive transmission allows targeting a specific packet in the ranging sequence, without affecting packets carrying data. Similarly, the attack packet does not contain any data field that could corrupt the content of the legitimate packet. The preamble is transmitted at low power so that it does not trigger a new receive event. Such an event would indeed lead to an error when the receiver determines the \ac{sts} quality and the presence of expected data fields. The \ac{sts} pulses are instead sent at higher power so that they overshadow the legitimate signal and produce noise that is misclassified as an early copy. Finally, both the power of preamble and \ac{sts} are adjusted based on the relative distance between devices. In particular, power is lowered to avoid jamming when the device that transmits the packet to overshadow is far away.  

\subsubsection{Selection of The Victim Packet(s)}
\label{subsec:packet-selection}
Typically, \ac{ds-twr} is used because it compensates for clock errors and
asymmetric reply times.
We have confirmed this in our analysis of many \ac{hrp} \ac{uwb} configurations.
As mentioned in the standard, distance is computed with the method proposed in~\cite{DBLP:conf/wpnc/NeirynckLM16}:
\begin{equation}
    \hat{d} = c \cdot \hat{T}_{prop} = c \cdot \frac{T_{round1}*T_{round2}-T_{reply1}*T_{reply2}}{T_{round1}+T_{round2}+T_{reply1}+T_{reply2}}
\end{equation}
For simplicity we can neglect non-idealities and consider that distance is measured as the average of the two rounds~\cite{6012487}:
\begin{equation}
    \hat{d} = c \cdot \hat{T}_{prop} = \frac{c}{4} \cdot (T_{round1}+T_{round2} - T_{reply1} - T_{reply2}) 
\end{equation}
Sometimes, a fourth message is used, likely for the transmission of additional data. 
If the ranging packets contain only preamble and \ac{sts} but no data, additional data packets are sent earlier and/or later. In any case, the attacker can configure the delay from reception of the first packet to attack either the second or the third packet of the \ac{ds-twr} sequence.
As shown in \autoref{fig:delay}, attacking the second packet corresponds to overshadowing a packet transmitted by the responder and received by the initiator, while attacking the third packet corresponds does the opposite. It is convenient for the attacker to be closer to the receiver to use less power for overshadowing, but it is not strictly necessary. 
It is worth noting that the choice between the second and third packet is not entirely symmetric. Attacking the third packet has the only effect of reducing the round time measured by the responder ($T_{round2}$) leading to: 
\begin{equation}\small
    \hat{d'} = \frac{c}{4} (T_{round1}+T_{round2}-\delta - T_{reply1} - T_{reply2}) \\ = \hat{d} - c \cdot \frac{\delta}{4} 
\end{equation}

Instead, attacking the second packet reduces both $T_{round1}$ (because the initiator receives the packet earlier) and $T_{round2}$ (because the initiator consequently replies earlier), leading to:
\begin{equation}\small
    \hat{d''} = \frac{c}{4} (T_{round1}-\delta+T_{round2}-\delta - T_{reply1} - T_{reply2})\\ = \hat{d} - c \cdot \frac{\delta}{2} 
\end{equation}
Clearly, by attacking the second packet, the attacker can obtain reductions that are twice as big as those obtained by attacking packet three.
The reduction $\delta$ is a random variable not in control of the attacker and bound by the maximum difference between \ac{los} and \ac{nlos} path accepted by the receiver.

As an alternative, the attacker can use devices to target both the second packet (near the initiator) and the third packet (near the responder). We can consider the two attacks as independent events. Therefore, the attacker will obtain reductions of $c\nicefrac{\delta}{4}$, $2c\nicefrac{\delta}{4}$, and $3c\nicefrac{\delta}{4}$, each with decreasing probability. We confirm this assumption in \autoref{sec:evaluation}.

\input{figures/setup.tikz}

    \begin{algorithm}[b!]\small
        \KwIn{\\
            \Indp \Indp
            Radio Channel\\
            STS Mode, Preamble, SFD configuration for Rx \\
            Preamble, SFD configuration for Tx\\
            STS length for Tx\\
            Preamble+SFD and STS power for Tx\\
            Delay from Rx to Tx
        }
        \Begin{
            ConfigureRx(RxConfig) \;
            ConfigureTx(TxConfig) \;
            \While{True}{
                WaitForRx() \;
                Wait(Delay) \;
                Tx() \;
            }
        }
        \caption{\label{algo:attack} Attack pseudo-code.}
    \end{algorithm}

\subsection{Implementation}
\label{ssec:attackimplementation}
We have presented a general approach to conduct distance-reduction attacks.
In principle, it can be implemented with any off-the-shelf \ac{hrp} \ac{uwb} \standardz compatible device that can be programmed to receive and transmit packets and that allows configuring individual power levels for each field.
In practice, we have implemented the attack using a Qorvo DWM3000EVB~\cite{DWM3000EVBQorvo}, controlled by a Nordic Semiconductor nRF52 DK~\cite{PCA10040Nordic}, for a total cost of around USD 65 only.
These devices can be easily programmed with open-source firmware~\cite{qorvofirmware}, they have limited size, and they can be powered by a portable USB battery. The pseudo-code of the attack is shown in \algoref{algo:attack}.
The delay can be configured to be a multiple of the reply time used by the
victims so that the attack signal is transmitted at the same time as
one of the following packets (\autoref{fig:delay}). The attacker can find
this and other reception parameters in an attack preparation phase.

For this, we have developed a sniffer and packet analyzer based on a Qorvo DWM3000EVB attached to an STM32 Nucleo-F429ZI.
Using Qorvo's SDK, we implement a fast \ac{uwb} receiver, which forwards frames over a USB connection to a host computer.
Here, packets are analyzed with a custom Wireshark dissector~\cite{wireshark} that also supports Apple's proprietary \ac{uwb} frame format.
Multiple packet analyzers with different configurations can be connected, which is required to observe complex ranging procedures.
The DWM3000EVB chip in our packet analyzer can receive timestamps with an accuracy of \SI{15.65}{\pico\second}\cite{DW3000USERMANUAL}.
These timestamps are recorded and forwarded to the Wireshark dissector.
As we will show later in \autoref{sec:evaluation}, accuracy in the order of \SI{}{\micro\second} is sufficient to run the attack.
Since most protocols that are using \ac{uwb} today are closed source, there is no option to analyze the protocols for potential privacy and security issues thoroughly. Besides ranging frames, the \ac{uwb} packet analyzer also receives data frames. 
This allows us to inspect if any private data, static identifiers, key material, or similar is shared over \ac{uwb}. 
Apple does not use the \standard MAC frame format, e.g., for iPhone to iPhone and iPhone to HomePod ranging. We implemented a Wireshark dissector that allows inspecting the parts of it that are not encrypted. 

\subsection{Application to Real \ac{hrp} \ac{uwb} Chips}
\label{ssec:attacksetup}
We successfully applied our distance-reduction attack against Apple U1 chips deployed in different products (iPhone, AirTag, HomePod) and interoperated with chips from other vendors (NXP SR040, NXP SR150, and Qorvo DWM3000).

\autoref{fig:iphone-plus-sr150} shows a concrete example. One iPhone 11 Pro (Apple U1) is placed at \SI{8}{\meter} distance from an NXP SR150 in line of sight.
The two devices exchange a total of 6 messages, where 3 are used for \ac{ds-twr}. The iPhone is the initiator (and victim) and the NXP SR150 is the responder. 
A Qorvo DWM3000EVB acts as an attacker placed at around \SI{30}{\centi\meter} from the victim iPhone. By hitting the second message of the \ac{ds-twr} sequence, the attacker 
causes distance reductions of up to \SI{10}{\meter}. The application running on the iPhone 
shows \SI{8}{\meter} when the attack is off and \SI{0}{\meter} during a successful reduction.

\autoref{fig:iphone-plus-iphone} shows another example of an attack targeting ranging between
two identical iPhones. In this case, the total number of messages is 4, but the attack is similar. 
By targeting the second packet of the \ac{ds-twr} sequence, the attacker causes reductions from \SI{10}{\meter} to less than \SI{2}{\meter} in the raw measurements plotted on the laptop.


\section{Experimental Evaluation}
\label{sec:evaluation}
In this section, we demonstrate the feasibility of our attack, show the number of distance reductions possible, and determine the success rate.

\subsection{Setup}
We ran the attacks in an indoor \ac{los} environment with two victim devices placed at various distances between \SI{5}{\meter} and \SI{15}{\meter} with antennas facing each other. This setup results in a relatively good baseline signal quality with a small ranging error (normally \SI{10}{\centi\meter} to \SI{20}{\centi\meter}) when the attack is turned off. 
We chose this setup to avoid measurement noise due to channel (e.g., excess paths) that would otherwise distort the outcome.
We evaluated the following device combinations: iPhone--iPhone (Nearby Interaction)\cite{apple-nearbyinteraction}, iPhone--AirTag (FindMy ranging)\cite{apple-airtag-ranging}, iPhone--HomePod (Handoff music)\cite{apple-homepod-handoff-audio}, iPhone--NXP and iPhone--Qorvo (compatibility mode)~\cite{apple-nearby-accessory}.

The attacker places either one or two Qorvo DWM3000EVB in ca. \SI{30}{\centi\meter} proximity to one or both ranging devices. The adversarial transceivers perform a reactive attack as introduced in \autoref{sec:attack}, i.e., they are programmed to detect the initial frame of the ranging exchange and then overshadow preamble and STS of one or two subsequent frames.
It is important to note that, while the overall success rate of the attack and the maximum distance reduction increases when two devices are employed, the result of the ranging procedure are synchronized among the devices, i.e., both legitimate devices eventually report the same measurement time series.

\begin{figure*}
\centering
\includegraphics[width=0.7\linewidth]{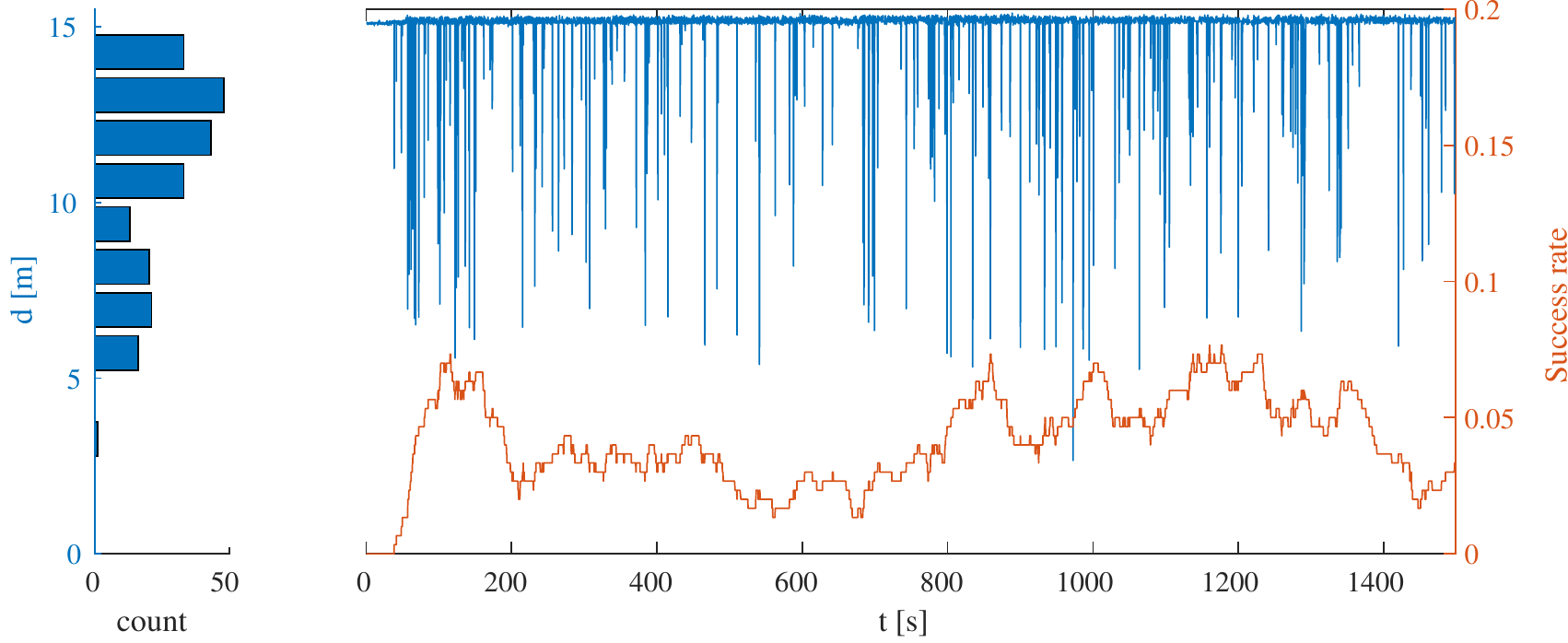}
\caption{A \SI{25}{\minute} trace of raw distance measurements (iPhone--iPhone, \SI{15}{\meter}) under attack with two devices. The left part is a histogram of those measurements that report reduced distances. The overall rate of successful distance reductions is \SI{4.08}{\percent}.
\label{fig:2qorvo_trace}}
\end{figure*}

\begin{figure}
\centering
\includegraphics[width=0.7\linewidth]{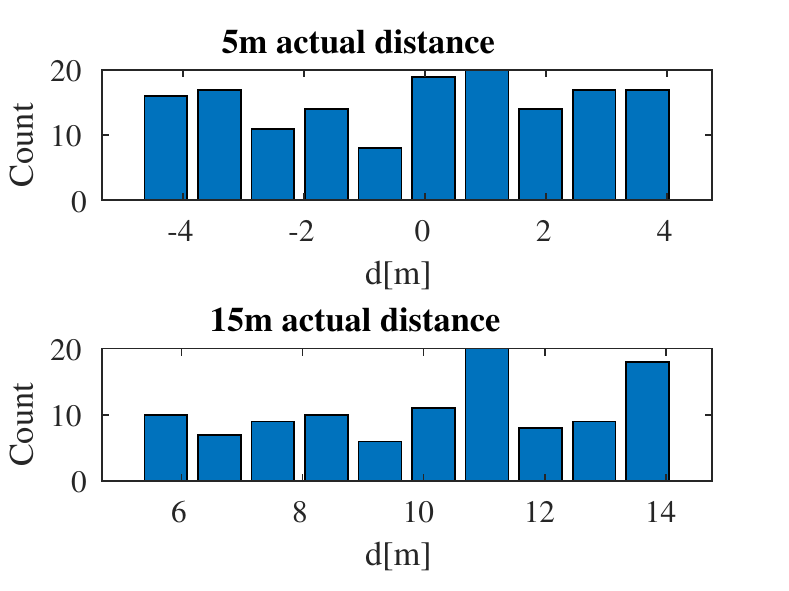}
\caption{Distribution of reduced distance reports for the iPhone--iPhone (\SI{5}{\meter} and \SI{15}{\meter}) setup, attacked with single device over a \SI{15}{\minute} observation period. The overall rate of successful distance reductions is ca. \SI{2.2}{\percent} in both cases.}
\label{fig:twodist_hist}
\end{figure}

\subsection{Retrieving Raw Distance Measurements}

\ac{uwb}-based key solutions only need to determine if a distance is below or above a threshold.
Thus, many applications do not display detailed distance information in the user interface.
In contrast, we need precise distance measurement results without aggregation to evaluate the success rates of attacks.
In the case of the Apple UWB implementation, the U1 chip reports the raw measurements to iOS drivers, which log them.
Viewing these measurement logs requires the \emph{Location Services} and \emph{AirTag} debug profile, which can be installed on any iPhone without jailbreak~\cite{apple-logs}.
Then, detailed measurement information appears in the logs, including the distance.
\begin{lstlisting}
@\textbf{nearbyd}@ #me,MeasEngMetricsCalculator::checkCirMetrics: AOA CycleIdx 1497 RangeMsmt 3.27164 machAbsTime @\textcolor{gray}{...}@
@\textbf{nearbyd}@ #sp,[Solution Provider]   @\textcolor{darkblue}{r1 range: 3.272 m}@
\end{lstlisting}

iOS will forward the raw measurement to the corresponding daemon or Nearby Interaction framework in most measurement modes.
For example, attacking a single distance measurement between an iPhone and a HomePod mini is sufficient
to show the HomePod's music playback menu on the iPhone immediately. We observe the same behavior in the compatibility mode, which
is used for third-party integration like car keys. For example, \autoref{fig:iphone-plus-sr150} shows a reduction to \SI{0}{\meter} visible on the screen.

Apple noticed that \ac{uwb} distance measurements are not always reliable.
When using the Nearby Interaction framework with the example \emph{Peekaboo} application~\cite{apple-nearbyinteraction},
the current measurement is only published if it does not deviate more than \SI{1}{\meter} from the median of the last \num{11} distance measurements.
This filter is not applied when the phone interacts with third-party devices.
We manually identify this boundary by replacing distance values reported by the U1 chip on an iPhone with \textsc{f\reflectbox{r}ida}~\cite{frida}, similar to previously published hooks~\cite{u1-bh}. iOS only discards measurements on the application layer,
thereby hiding them from curious developers.
In the following, we use the raw U1 chip measurements provided by the logs of an unmodified iPhone to get comparable results, irrespective of opaque application-layer filters.

\subsection{Results}
We quantify the success of the attack as the relative rate of ranging measurements (as read from the iOS logs) indicating a distance shorter than the baseline, averaged over an observation interval of at least \SI{15}{\minute}. To separate benign measurement non-idealities from actual reductions, we only count measurements lower than two times the maximum benign (negative) deviation during a \SI{100}{\second} interval before running the attack.

For different device combinations, our attack causes distance reductions between ca. \SIrange{2}{12}{\meter} with success rates in the range of \SIrange{2}{4}{\percent}. An overview is provided in \autoref{table:attack-scenarios}. Some of the differences in success rates, i.e., those between \SI{2}{\percent} and \SI{4}{\percent}, can be explained by the fact that either only one or two packets of the ranging procedure are attacked. This means, for a given success rate per individual ToA measurement (i.e., by packet), we increase the chances that at least one ToA measurement of the ranging exchange is successfully reduced by targeting both the second and third packet.
To exemplify the cumulative effect of multiple attacking devices on the overall reduction, \autoref{fig:2qorvo_trace} shows the entire time series of measurements (iPhone--iPhone) over a \SI{25}{\minute} observation interval with one attack device placed at each end. Due to the attack success rate changing over time, we also display the instantaneous success rate using a sliding window over \num{300} consecutive measurements. The attack results in an overall rate of reduced measurements of \SI{4.1}{\percent}, whereas the rolling average over \num{300} consecutive packets can get as high as \SI{7.7}{\percent}. The distribution of distance reductions is biased towards reductions $\leq$\SI{5}{\meter} because either of the devices, i.e., the one targeting the second packet and the one targeting the third packet, can cause those. In contrast, the device replying to the initiator (i.e., transmitting over the second packet) can solely have an effect up to \SI{10}{\meter}. This observation is in line with the analysis provided in \autoref{sec:working-principle}. The longest reduction observed over this interval is over \SI{12.35}{\meter}, caused by successful reductions on both packets attacked during the same ranging procedure. Assuming independence of the effects on either side, this shows that these additive reductions (exceeding \SI{10}{\meter}), while orders of magnitude less likely, are still frequent enough to occur within a realistic time window (\SI{25}{\minute}). Similarly, in a configuration where only the responder is vulnerable, distance reductions are limited by ca. \SI{5}{\meter}, because only the ToA of the third packet can be targeted. An example for this is the combination of iPhone and NXP SR040, since NXP SR040 can only be configured as initiator.

The range of possible relative distance reductions does not depend on the actual distance of the ranging devices, and the U1 chip even reports negative distances in case the distance reduction exceeds the nominal distance. \autoref{fig:twodist_hist} highlights this, showing the distribution of reduced distance reports in the iPhone--iPhone setup with one attack device over two different distances, \SI{5}{\meter} and \SI{15}{\meter}, over a \SI{15}{\minute} observation period. It becomes evident that the relative reduction is, irrespective of the nominal distance, bounded by \SI{10}{\meter}.

\paragraph{Jamming}
Even though organizations claim that \ac{uwb} is \emph{immune to jamming}\cite{fira-uwb-technology-comparison}, we saw that it is perfectly possible to disturb ranging measurements through jamming. When setting the transmission power to a high level and the transmission time to match with an expected ranging frame the receiver will not be able to receive the frame. This is likely caused by disturbing the \ac{sts}.

\subsection{Parameter Tuning} \label{sec:parameter_tuning}
The timings and power level of the adversarial signal have to be matched to the legitimate signal. To run a successful attack, a time delay that matches the actual frame has to be accurate in the order of \SI{5}{\micro\second} to have success rate of above \SI{1}{percent}. 
Whereas the timing is relatively static, the power level depends on the baseline signal quality. This means the attack power needs to be adjusted depending on the path loss, i.e., the distance between the legitimate devices. E.g., for victim devices at a nominal distance of 15m, we adjusted the gain parameters to a fraction of the maximum possible value on the Qorvo, ca. \nicefrac{1}{4} for the preamble and \nicefrac{1}{3} for the STS.

\subsection{Device Pairings}
In \autoref{table:attack-scenarios} we show the results of performing the attack against different pairs of devices. In most cases, the iPhone has been the main victim, since its implementation seems to be most affected by this vulnerability. 
Our results have shown that one vulnerable device results in a distance reduction for both devices. This issue cannot be mitigated on one end only, since every \ac{uwb} ranging algorithm requires both devices to report round-trip time $T_{round}$ and reply delay $T_{reply}$ to the other devices. This means that a user has to trust both devices, which can only be achieved through independent certification, including a review of the algorithms. 

Additionally, we see that it is irrelevant with which device the iPhone performs ranging. Every device combination is vulnerable to distance reduction with a good success rate.

\begin{table*}[]
\footnotesize
\centering
\renewcommand{\arraystretch}{1.3}
\caption{Overview of attack scenarios and results.}
\vspace{-1em}
\label{table:attack-scenarios}
\begin{tabularx}{0.8\linewidth}{lllrr}
\toprule
Scenario            & Primary Victim (V1) & Secondary Victim (V2) & Maximum Reduction & Success Rate \\ 
\midrule
Handoff Music & iPhone (Apple U1) & HomePod mini (Apple U1) & \SI{9.01}{\meter} & \SI{2.10}{\percent} \\
Nearby Interaction  & iPhone (Apple U1) & iPhone (Apple U1) & \SI{12.45}{\meter} & \SI{4.08}{\percent} \\
AirTag               & AirTag (Apple U1) & iPhone (Apple U1) & \SI{9.09}{\meter} & \SI{4.25}{\percent} \\
NXP Initiator   & iPhone (Apple U1) & Tag (NXP SR040) & \SI{4.80}{\meter} & \SI{1.87}{\percent} \\
NXP Responder   & iPhone (Apple U1) & Tag (NXP SR150) & \SI{9.68}{\meter} & \SI{2.15}{\percent} \\
Qorvo & iPhone (Apple U1) & Tag (Qorvo DWM3000) & \SI{8.13}{\meter} & \SI{3.09}{\percent} \\ 
\bottomrule
\end{tabularx}
\end{table*}

\subsubsection{Confirmation of Results by Binary Analysis}

Our attacks work without knowing \ac{uwb} chip implementation details.
Nonetheless, binary analysis of the \ac{uwb} implementations helps understanding why attacks are feasible.

All \ac{uwb} chips analyzed in this paper are split into a main application and a low-level \acf{dsp}.
The \ac{dsp} can be instrumented over a serial interface. The NXP and Qorvo chip have a documented
\ac{api} for this interface. However, the \ac{dsp} itself is inaccessible on these platforms.
Qorvo does not allow reprogramming the \ac{dsp} to the best of our knowledge. NXP ships firmware
files for the \ac{dsp}, but they are encrypted and signed, even in the development kit, which prevents analysis.

In contrast, Apple's U1 \ac{dsp} firmware is part of the software updates for all \ac{uwb}-enabled
devices. It ships in a proprietary \texttt{ftab} format~\cite{u1-bh}, and the firmware is not encrypted. However, it is a bare metal firmware without any symbols.  It contains a few strings, including assertions about the distance measurement.
One of these strings refers to the backsearch window and is part of a function used for \ac{sts} correlation:
\begin{lstlisting}[language=C]
(inp->sum_window_right - inp->sum_window_left + 1) <= 16
\end{lstlisting}

Since the backsearch window sampling rate is unknown, we cannot
calculate the maximum possible distance reduction.


\section{Discussion}
\label{sec:discussion}
\subsection{Strengths and Limitations}
We proposed a practical attack that achieves distance reduction of several meters using only a simple and inexpensive off-the-shelf device.

Its practicality makes this attack particularly relevant for security-critical application that are gaining more market traction, for example, \ac{pkes} systems in cars.

The main limitation of the attack is the little control on the amount of distance reduction caused by selective overshadowing. The maximum reduction is set by the maximum difference between \ac{nlos} and \ac{los} accepted by the victim, and how many packets in the \ac{ds-twr} sequence are targeted.
In the case of Apple U1, this results in attacks that reduce distance by a maximum of \SI{5}{\meter} when attacking the third packet of \ac{ds-twr}, \SI{10}{\meter} when attacking the second, and \SI{15}{\meter} when attacking both. Nevertheless, it is worth noting that the attacker is generally not interested in steering distance precisely. For example, to attack an access control system, it is enough for the attacker to cause a reduction below the threshold that grants access within a reasonable time frame.

A related limitation is that the attacker cannot control precisely which ranging sequence is affected by a reduction.
Thus, the victim could try to identify outliers as distance reduction attack.
However, in a practical setting, the users move and have their car key in a pocket, leading to similar outliers.

\subsection{Reflections on \ac{hrp} \ac{uwb} Security}
Our results are a clear call for research to improve the security of HRP. We have shown that current security properties hinge entirely on the quality of proprietary algorithms and black-box implementations and that mere compliance with IEEE 802.15.4z does not protect systems against distance reduction attacks. We argue that security should not be a distinguishing feature of individual implementations but publicly available and verifiable. We are convinced that a secure standard that withstands the scrutiny of the research community also benefits device vendors, as they can rely on its correctness guarantees and focus entirely on implementation challenges. Consequently, it is important to direct future work towards the development of a secure and open algorithm for first path detection, which can be integrated into an upcoming standard.
The long \ac{sts} of \ac{hrp} \ac{uwb} intuitively suggests a high level of security. However, the attack success rate achieved in practice by attacking the leading edge detection is orders of magnitude higher than the probability of guessing the \ac{sts}. The main takeaway of our results is that there is no guarantee that the length of the \ac{sts} represents the actual security level of a system based on \ac{hrp} \ac{uwb}.

\subsection{Countermeasures}
The necessity to distinguish legitimate early copies of the signal from the attacker's induced noise creates a tension between \ac{hrp} \ac{uwb} security and performance.
Our results indicate that Apple's current U1 implementation does not perform any noticeably advanced checks on the \ac{sts}.
However, the chip does perform some basic checks, since we cannot simply achieve \SI{100}{\percent} success rate by transmitting packets in advance.
The receiver must apply advanced statistics on the incoming signal to detect attacks while still performing well in challenging \ac{nlos} conditions. 
For example, the receiver could compare the consistency between the channel response of the early low-power copy and the main copy. The \ac{sts} quality, which represents the similarity of the incoming signal with the expected one, should be checked independently for both early and late copies. In other words, the receiver should not assume that an early copy is acceptable just because another valid late copy appears just afterwards.
Such countermeasures require increasing the complexity of the receiver,
which might not be feasible for battery-powered devices like the AirTags.

Reducing the maximum accepted difference between \ac{los} and \ac{nlos} copies would reduce the maximum reduction that an attacker can achieve, but it would also limit the use of the product in realistic scenarios, i.e., a car key is a pocket.

More countermeasures can be applied at the upper layers, for example, detecting reductions as outliers. 
However, real-time applications do not have a margin for accumulating more than a few measurements before reacting based on the measurement value.

\subsection{Future Work}

On the theoretical side, our results are a call for researching whether it would be possible to assure the \ac{hrp} \ac{uwb} security level. This is challenging because its security is intertwined with proprietary algorithms and design choices.
On the practical side, the analysis and attack phases of our attack could be combined in a feedback loop.
For example, the power of the attack packet could be automatically chosen based on some observations at reception. A chip such as the Qorvo DWM3000EVB offers many reception diagnostics. For example, when receiving the packets sent by the victims, the attacker could estimate the quality of their preamble, obtain a rough estimate of its relative distance from the victims, and adjust the power used for overshadowing. Similarly, the attacker could measure the rate of packets exchanged by the victims and lower its power if it detects a lower rate due to jamming. Other attack parameters could be adaptively configured in a similar automated fashion.

\section{Related Work}
\label{sec:related}
The \ac{uwb} \standard standard is described in~\cite{9144691,9179124}. Chips following the \ac{hrp} mode of the standard have been implemented by several vendors, such as Apple (U1)~\cite{apple-uwb}, NXP (SR040, SR150, SR100T)~\cite{nxp-factsheet}, and Qorvo (DWM3000)~\cite{DWM3000EVBQorvo}. Chips implementing the standard \ac{lrp} mode have been implemented by Microchip (ATA8352, ATA8350)~\cite{microchip} and Renesas~\cite{renesas}. To the best of our knowledge, these \ac{lrp} mode chips are not available in consumer electronic devices.

The first implementation-independent security evaluation of \ac{hrp} \ac{uwb} at the physical layer has been conducted in~\cite{DBLP:conf/wisec/SinghRZLC21}. That work proposed two attacks on HRP, derived from the Cicada attack~\cite{5616900,DBLP:journals/twc/PoturalskiFPHB12}, and shows in simulations that even conservative receiver implementations could be susceptible to distance reductions. In contrast to our paper, the authors neither conducted experiments with real \ac{uwb} chips, nor prove that attacks are practical with off-the-shelf hardware. Furthermore, they did not consider other aspects of the \ac{uwb} ranging protocols, i.e., the sequence of messages, significance of different message fields, or their power.

Further research on \ac{uwb} ranging has been done in~\cite{DBLP:conf/ndss/SinghLC19}. This work proposes improvements to \ac{lrp} that aim at securely extending the range of the \ac{lrp} through pulse interleaving. 

Previously documented attacks against \ac{uwb}~\cite{DBLP:conf/wisec/FluryPPHB10}, which applied to earlier standards (\standarda), cannot be used against \ac{hrp} because of the high frequency of the pulses.

The Apple U1 chip and secure ranging in iOS have been studied recently in~\cite{u1-bh}, with a focus on the overall software architecture rather than physical-layer aspects. After the release of AirTags, the hardware hacking community has discovered that its main firmware can be easily extracted and modified by glitching~\cite{airtag-colin,airtag-defcon}. Although the mentioned work provides an interesting overview on Apple's usage of \ac{hrp} \ac{uwb} in its products, it has not analyzed physical-layer attacks or features of the firmware that have an impact on the security of the physical layer (e.g., the code of the digital signal processor or the logic that decides whether to accept early peaks in the back-search window).

Other studies~\cite{DBLP:journals/popets/HeinrichSKH21} focused on the security of the Apple FindMy network that allows locating devices including AirTags. It is possible to add custom \ac{ble} devices to the FindMy network~\cite{DBLP:conf/wisec/HeinrichSH21} or to leverage FindMy to upload data from devices without Internet connection~\cite{sendmy}. These works focus on \ac{ble} and the architecture of FindMy and are mostly unrelated to the \ac{uwb} technology that some of the devices deploy.

The security of time of arrival measurements has been formalized in the form of Message Time of Arrival Codes (MTAC) in~\cite{DBLP:conf/sp/LeuSRPC20}.


\section{Conclusion}
\label{sec:conclusion}
We demonstrated for the first time a practical distance reduction attack against \ac{hrp} \ac{uwb} (\standardz) secure ranging, implemented in Apple U1 chips and widely deployed in Apple products. We demonstrate that the impact reaches beyond the Apple ecosystem, showing attacks when ranging is performed between an Apple U1 chip in an iPhone and development kits with chips by NXP and Qorvo. Distance reduction is a considerable concern in many applications, from access control (e.g., opening cars, doors) to mobile payments and indoor positioning for industrial plants. Our attack is practical, and it can be implemented with a cheap off-the-shelf device.
Our results raise the awareness on the pitfalls of \ac{hrp} \ac{uwb} technology. On the one hand, \ac{hrp} \ac{uwb} promises a nominally high security level based on a cryptographically secure STS sequence that cannot be guessed by an attacker. On the other hand, the actual security level depends on obscure design choices at the receiver. No independent experimental evaluation and certification framework exists either. 
Our results show that distance-reduction attacks are practical.
To improve the state of \ac{hrp} \ac{uwb} security, we have proposed and discussed several countermeasures.



 \section*{Availability}
Source code for the \ac{uwb} sniffer, packet analyzer, and Wireshark dissector will be made available.

%
%
%
\begin{acronym}
\acro{pkes}[PKES]{Passive Keyless Entry and Start}
\acro{uwb}[UWB]{Ultra-Wide Band}
\acro{hrp}[HRP]{High-Rate Pulse Repetition Frequency}
\acro{lrp}[LRP]{Low-Rate Pulse Repetition Frequency}
\acro{twr}[TWR]{Two-Way Ranging}
\acro{ds-twr}[DS-TWR]{Double-Sided Two-Way Ranging}
\acro{ss-twr}[SS-TWR]{Single-Sided Two-Way Ranging}
\acro{tdoa}[TDOA]{Time Difference of Arrival}
\acro{prf}[PRF]{Pulse-Repetition Frequency}
\acro{psd}[PSD]{Power Spectral Density}
\acro{toa}[ToA]{Time-of-Arrival}
\acro{tof}[ToF]{Time-of-Flight}
\acro{los}[LoS]{Line-of-Sight}
\acro{nlos}[NLoS]{Non-Line-of-Sight}
\acro{sts}[STS]{Scrambled Timestamp Sequence}
\acro{sfd}[SFD]{Start-of-Frame Delimiter}
\acro{mitm}[MitM]{Machine-in-the-Middle}
\acro{ni}[NI]{Nearby Interaction}
\acro{ble}[BLE]{Bluetooth Low Energy}
\acro{edlc}[ED/LC]{Early Detect/Late Commit}
\acro{aoa}[AoA]{Angle of Arrival}
\acro{toa}[ToA]{Time of Arrival}
\acro{dsp}[DSP]{Digital Signal Processor}
\acro{api}[API]{Application Programming Interface}
\acro{cir}[CIR]{Channel Impulse Response}
\acro{sts}[STS]{Scrambled Timestamp Sequence}
\acro{mfi}[MFi]{Made for Apple}
\acro{bpsk}[BPSK]{Binary Phase Shift Keying}
\acro{bpm}[BPM]{Burst-Position Modulation}
\end{acronym}

\smaller
\bibliographystyle{plain}
\bibliography{paper.bib}

\end{document}